Foreword: the paper below describes a concrete tool for protein physics. The mathematical basis for this approach comes from topology and differential geometry. The seminal paper in this field is Chern, S.-S.; Simons, J. (1974). "Characteristic forms and geometric invariants". *Annals of Mathematics* **99** (1): 48–69. This paper is couched in the language of pure mathematics, which has subsequently been translated into the language of quantum field theory, for instance http://en.wikipedia.org/wiki/Chern-Simons_theory. The vector bundles discussed by Chern-Simons are topologically similar to self-organized glass and polymer networks. The boundary problems they discuss disappear from standard elementary particle theory because of the large gap in particle masses. They also disappear from the constraint theory of network glasses because there are often similar gaps in hierarchical valence force fields, especially in networks where the nodes (atoms) have similar sizes, for instance Ge-As-Se glass alloys (Phil. Mag. **85** (2005) 3823). In proteins the boundary problems remain, but are solved by a different method, described below, based on hydrophobic compaction of protein globules.

# Self-Organized Criticality: A Magic Wand for Protein Physics

## J. C. Phillips

Dept. of Physics and Astronomy, Rutgers University, Piscataway, N. J., 08854

## Abstract

Self-organized criticality (SOC) is a popular concept that has been the subject of more than 3000 articles in the last 25 years. Here we show that SOC may enable theory to connect standard Web-based (BLAST) short-range amino acid (aa) similarities to long-range aa roughening form factors that accurately describe evolutionary trends in water-membrane protein interactions. Our method utilizes a hydropathic aa metric based on 5526 protein segments and thereby encapsulates differential geometrical features of the Protein Data Bank. It easily organizes small aa sequence differences between humans and proximate species. For rhodopsin, the most studied transmembrane signaling protein associated with night vision, it shows that short- and long-range aa sequence properties correlate with 96% success for humans, monkeys, cats, mice and rabbits. Proper application of SOC promises unprecedented simplifications of exponentially complex protein sequence-structure-function problems, both conceptual and practical.

It is widely believed that magic wands occur only in fairy tales. Proteins are the canonical example of exponentially complex networks: there are $10^{385}$ possible aa sequences for a typical protein with 350 aa side groups on its peptide chain. Suppose a method exists for organizing the properties of homologous members of protein families, and arranging them hierarchically to exhibit evolutionary refinements. Suppose further that this method contains no adjustable parameters and could be employed on a PC using only EXCEL macros. We show here that such a method exists, and it yields a cornucopia of unexpectedly precise results, unobtainable by other methods, justifying calling it a molecular magic wand.

SOC originated in the context of sand piles[1]; the critical exponents of its transport properties are known for various models[2]. There are at least five reasons optimized proteins are more complex than marginally stable sand piles: (I) their basic peptide units are polymerized, (II) each unit has an aa side group, chosen from a menu of 20 candidates (the sand grains - polymer beads come in 20 colors), (III) the protein chains have been compacted hydrophobically into space-filling globules, (IV) proteins can live and function, and (V) protein aa sequences and secondary structures (such as hydrogen-bonded helices, strands and loops) have been optimized by evolution.

Because of (III), hydropathic critical exponents $\psi(aa)$ can be identified that are specific to each aa, and the resulting hierarchical list broadly resembles ones obtained from aa transfer energies from water to organic solvents[3]. There is, however, an important difference: the dimensionless hydropathic critical exponents refer to the evolution of solvent accessible surface areas (SASA, for a 2 A water molecule) of protein segments of length 2N+1, only at long range with $4 \leq N \leq 17$, and thus are explicitly separated from short-range steric packing interactions $N < 4$. The latter short chain segments constitute a geometrical problem that at present is solved most often statistically by tools specific to given situations. Nevertheless, we have found that many trends in the properties of structurally homologous proteins correlate well with only the properties of the long-range SOC critical exponents $\Psi = \{\psi(aa)\}$ defined by

$$\text{dlog(SASA(aa))/dlogN} = -\psi(aa) \qquad (1)$$

Protein functionality is often determined by weak, long-range hydropathic interactions ($N \geq 4$), rather than strong, short-range steric ones ($N < 4$), and this is why SOC can be a magic wand for protein physics.

Suppose that the complex functional path followed by a protein $\mathscr{P}$ is dominated by a single configuration coordinate $\mathscr{C}$ based on an optimized convolution of the hydrophobicity $\Psi(\mathscr{P})$ derived from the protein amino acid sequence $\mathscr{S}$ Because of (V), it is then not necessary to carry out reductionist molecular dynamics simulations (MDS) on $\mathscr{P}$ in an aqueous environment (at present inaccurate and restricted to small $\mathscr{P}$ with $< 150$ aa)[4]; instead, many properties can be explored by studying $\mathscr{C}(\Psi(\mathscr{S}(\mathscr{C})))$ where $\{\mathscr{C}\}$ is a set of proteins $\mathscr{C}$ proximate to $\mathscr{P}$; such sets are easily constructed from $\mathscr{P}$ simply by using BLAST.

This program was previously applied to lysozyme $c$ (Hen Egg White), a remarkably robust 130 aa protein, whose peptide backbone $C_\alpha$ coordinates for chicken and human are structurally superposable to 0.65 A (nearly unchanged in 400 million years!), yet with 58 aa mutations[5]. There the dual metabolic and antibiotic properties are known for two proximate sets $\{\mathscr{C}\}$ (birds and placental mammals); these correlated well separately and together with profiles of $\mathscr{C} = <\psi 3>$, where $<\psi W>$ is $\psi$ averaged over a sliding window of length W.

Here we discuss a superfamily $\{\mathscr{C}\}$ of transmembrane proteins with 800 human members, of great pharmaceutical interest, and so well studied: the Guanine Protein Coupled Receptor (GPCR) superfamily, the largest family of proteins in the human genome. GPCR proteins have characteristic heptad structures, with seven long (25aa), predominantly helical, TransMembrane (TM) interior sections connected by exterior or surface ExtraCellular (EC) and CytoPlasmic (CP) loops[6]. Their amino acid sequences form the largest database for protein-membrane interactions, and these protein receptors perform a variety of signaling functions: rhodopsin (visual), adrenergic (stimulative), adenergic (metabolic), etc.

Given their known heptad structures, and the locations of their chemical receptors (retinal, adrenalin, etc.) between transmembrane (TM) interior sections[6], it is clear that functionality will depend on larger values of sliding window lengths W in GPCR proteins than in lysozyme, where W = 3 worked well. On this larger W scale one can use macroscopic concepts such as protein-membrane interfacial roughening, or interfacial water density fluctuations. To lowest order in hydropathic fluctuations, these will depend on $\mathscr{C} = <\psi W>$, where the dimensionless window length W can be optimized, and specifically on the overall variance (quadratic fluctuations from average), represented by the hydropathic roughening $\mathscr{A}(C(\mathscr{A}W)))$. The latter can be refined by calculating quadratic fluctuations using three separated average values, for the TM, EC and CP regions, denoted by $\mathscr{R}*(\mathscr{A})$.

Interfacial fluctuations should be highly sensitive to pressure. At high pressures, smoother sequences may be more flexible and functionally more effective and less subject to damage by pressure fluctuations. An elegant example that tests this idea is { $\mathscr{C}_{hp}$ ∕ river and deep sea (high pressure) lamprey rhodopsins, which are encoded by a single gene[7]. The aa of the two species differ at 29 out of 353 sites, and three of these have been identified as responsible for causing a blue shift in the rhodopsin absorption spectra for adaptation to the blue-green photic environment in deep water[7]. This leaves 26 aa replacements to be explained. Structurally 20 out of 171 differences are located in TM regions, and 9 out of 182 in EC and CP loop sites. The predominance of TM substitutions by a factor of 2.2 is understandable, as the increased deep-water pressure constrains internal pore-confined TM motion more than surface loop motion.

When we compare BLAST similarity and $\mathscr{D}$*(W) for the two rhodopsin adaptations { $\mathscr{C}_{hp}$ ∕ with human rhodopsin, quite a different picture emerges. For W ≤ 25 (TM length L = 25 or shorter), the differences are small, but at sliding window W = 47 (2L), they are large (Table I). The W = 47 sliding window profiles (Fig. 1) show that these deviations are concentrated in a few cytoplasmic secondary structures (100-150 (TM2-CP2-TM3) and 290-end (TM7-CP4)). What is most striking is that sum of the river lamprey rhodopsin's $\mathscr{D}$*(47) excess fluctuations over the deep sea lamprey rhodopsin's exceeds the reverse sum by a factor of 3.5, a factor 60% larger than the concentration of mutations in structural TM compared to EC and CP loops. Given the linear limitations of sliding profiling for fixed W = 47, this is persuasive evidence that the main function of the 29 deep sea lamprey rhodopsin mutations is to smooth very long wave length hydropathic fluctuations.

It is all very well to discuss large differences between river and marine lampreys, but the pharmaceutical industry is much more interested in the small differences between humans, monkeys, cats, mice and rabbits. For non-marine mammals, human rhodopsin (348 aa) can be used as an absolute benchmark, and the short-range BLAST similarity of the rhodopsin of these five proximate species {$\mathcal{G}_h$} to human rhodopsin is listed in Table II.  The evolutionary hierarchy given by BLAST is pretty much as expected: human, monkey, (rabbit, cat), mouse. The correlation coefficients C = |R| of the short-range BLAST scores with $\mathcal{R}$*(W) are impressive, as even for W = 1, C = 0.54 (compare to conventional MDS "folding"[4], limited at present to less than 50% success for proteins smaller than 150 aa), while for  nearest neighbors, W = 3, already C = 0.86 (excellent!), but the optimal value occurs at W ~ 25 (1 TM length), and here C = 0.96, a remarkable success which is possible only because of SOC.

One could even say that these five species {$\mathcal{G}_h$} define a critically optimized rhodopsin subfamily, larger than primates, but smaller than mammals.  Careful examination of Table II shows that dog begins to diverge from the subfamily with low similarity at both short range (BLAST) and longest range roughness $\mathcal{R}$*(47), while  $\mathcal{R}$*(W)   maintains the subfamily correlation for the broad midrange $3 \leq W \leq 25$.

We can explore these unexpected correlations (comparing alphabetical short-range BLAST similarities to numerical long-range $\mathcal{R}$*(W) resembles comparing apples to oranges) in two ways: (A) use the MZ scale[3], but compute the roughness as a simple variance with a common protein-wide average, and (B) continue to separate EC, CP and TM regions, but use the short-range water-organic transference hydropathic KD scale[3,5] instead of the long-range MZ  SOC scale.  The results (Fig. 2) show that with (A) the maximum C = 0.96 seen near W = 25 (TM length) disappears, while C still remains large > 0.93±0.02, while (B) the KD transference scale shifts the peak in C to  W ~ 10, a medium-range length which may reflect the harmonic average of short-range transference interactions ( W~ 4) and the intrinsic TM length W ~ 25.  In practice, since the short-range interactions are already well handled by BLAST-based libraries for specific protein families, the advantages of using $\mathscr{R}^*$(W) [which here attains the remarkable maximum of $\mathscr{R}^*$(25) = 0.96, with only 4% rounding errors] to treat long-range interactions separately are obvious.

It would appear that the very high rhodopsin correlation coefficient $\mathscr{R}^*$(25) = 0.96 must be accidental, since there seems to be a gap between the BLAST values (which would appear to be similar to $\mathscr{R}^*$(1)) and $\mathscr{R}^*$(9). In fact, studies of other opsins (not included here), specifically the red cone opsin, show that this high value $\mathscr{R}^*$(25) = 0.96 occurs because rhodopsin itself is found in cylindrical (quasi-one-dimensional) rod cells that peripherally support and stabilize central trichromatic cone receptor arrays. Opsin activated states involve tilted transmembrane segments[8], and such tilts may be minimized both by rod mosaic configurations, and by a specially stable matching of short-and long-range rhodopsin aa sequences, which reaches its evolutionary maximum in humans.

While these physiological results are striking, the real value of the results shown in the Tables is much greater. It is clear that the critical long-range hydropathic hierarchies are functionally much more successful than the standard short-range BLAST multiple alignment similarities, and that they can be combined with the latter to analyze adaptive plasticity and protein network stresses in a wide range of contexts[9-11]. The ability to analyze rhodopsin ultra-proximate interspecies differences has been demonstrated here for two cases: deep sea marine environment, and five (or six) human-proximate maximally evolved mammals, while many other cases are analyzed elsewhere[7].

The ability to analyze human-mouse-rabbit functional hydro-stress induced differences without adjustable parameters could be most useful in the context of engineering humanized mouse- or rabbit-derived monoclonal Antibodies (mAbs), which are receiving much pharmaceutical attention[12]. For immunoglobulin structures, one would also use three regions (heavy chain, light chain, and J), and separated long-range hydropathic analysis should be a useful supplement to short length (L = 9) family-specific libraries[11].

Overall it appears that the hydropathic form factors $\mathscr{A}(W)$ and $\mathscr{A}^*(W)$ readily lend themselves to recognizing qualitative trends associated with long-range interactions and analyzing them quantitatively. It is important to realize that although the present abstract hydropathic sequential analysis based on SOC is based on structural homologies, it delivers highly detailed, physiologically valuable information that is unobtainable by structural studies alone, now or in the foreseeable future. It represents the sequence-specific realization of general ideas about the importance of short- and long-range transmembrane structure interactions, hitherto treated only by a coarse-grained model[13]. Similarly it demonstrates in the specific context of globular hydrophobic protein folding the importance of competing short- and long-range forces to fill space glassily without forming ordered structures[14]. Studies of forces observed between hydrophobic surfaces suggest that the packing effects below N = 4 and W = 9 discussed here are associated with water structuring effects[15]. There are also very limited experiments to determine short-range roughness, based on proxy measures of fitness, such as catalytic activity, protein stability, or drug resistance[16]. Finally, it should be mentioned that macroscopic roughness and hydrophobicity were connected in an earlier work[17] which did not realize that proteins actually are archetypical examples of hydrophobically self-organized criticality[3].

# Table Captions

Table I.  Although the deep sea lamprey rhodopsin appears to be little different from the river lamprey rhodopsin as regards short-range BLAST sequence similarity, it is smoother than human rhodopsin at W = 25 ~ 1 TM length, and almost as smooth as human rhodopsin at W = 47 ~ 2 TM lengths, where the river lamprey rhodopsin is almost as rough as chicken rhodopsin (not shown).

Table II.  BLAST similarity scores (compared to human), relative roughness scores $\mathscr{R}$*(W) for W = 3, 9, 25 and 47 for five mammalian species.  Human roughness is smallest, and the interspecies differences increase with increasing W, with the best correlation to BLAST occurring at the transmembrane length W = 25.  Note that rabbit (prey, herbivore) is smoother than cat (predator, carnivore) for large W, although the BLAST scores are equal.  The quoted BLAST- $\mathscr{R}$*(W) correlation C does not include dog (see text).

# Figure Captions

Fig. 1.  MZ hydropathic profiles averaged over a W = 47 window.  Note the dramatic smoothing of the marine deep lamprey, especially in the 100-150 (TM2-CP2-TM3) and 290-end (TM7-CP4) regions.  Amino acid numbering as in Uniprot.

Fig. 2.  Rhodopsin correlation C of roughening $\mathscr{R}_{MZ}$(W) $\mathscr{R}_{MZ}$*(W) and $\mathscr{R}_{KD}$*(W) with BLAST similarity to human of five species (humans, monkeys, cats, mice and rabbits).

|  | Human | River lamprey | Deep sea lamprey |
|---|---|---|---|
| Uniprot | P08100 | Q90215 | Q90214 |
| BLAST/Human | 1.00 | 0.82 | 0.83 |
| $\mathscr{R}$*(25)/Human | 1.00 | 1.04 | 0.97 |
| $\mathscr{R}$*(47)/Human | 1.00 | 1.59 | 1.10 |

Table I.

|  | BLAST | $\mathscr{R}$*(3)/Human | $\mathscr{R}$*(9)/Human | $\mathscr{R}$*(25)/Human | $\mathscr{R}$*(47)/Human |
|---|---|---|---|---|---|
| Correl. (\|R\|) |  | 0.86 | 0.92 | 0.96 | 0.92 |
| Human | 717 | 1.00 | 1.00 | 1.00 | 1.00 |
| Monkey | 705 | 1.01 | 1.02 | 1.07 | 1.05 |
| Rabbit | 701 | 1.03 | 1.04 | 1.07 | 1.09 |
| Cat | 701 | 1.04 | 1.06 | 1.11 | 1.16 |
| Mouse | 692 | 1.04 | 1.11 | 1.14 | 1.23 |
| [Dog] | 675 | 1.06 | 1.13 | 1.16 | 1.20 |

Table II

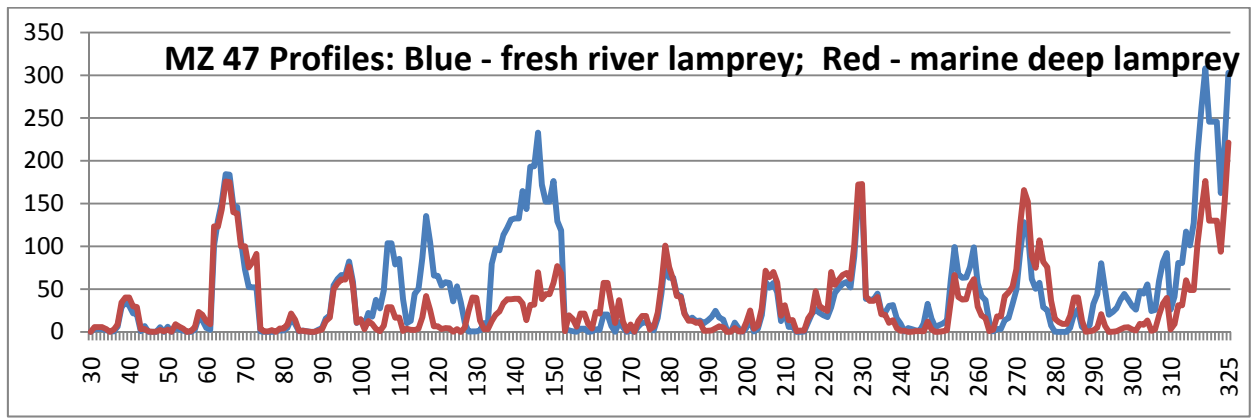

Fig. 1.

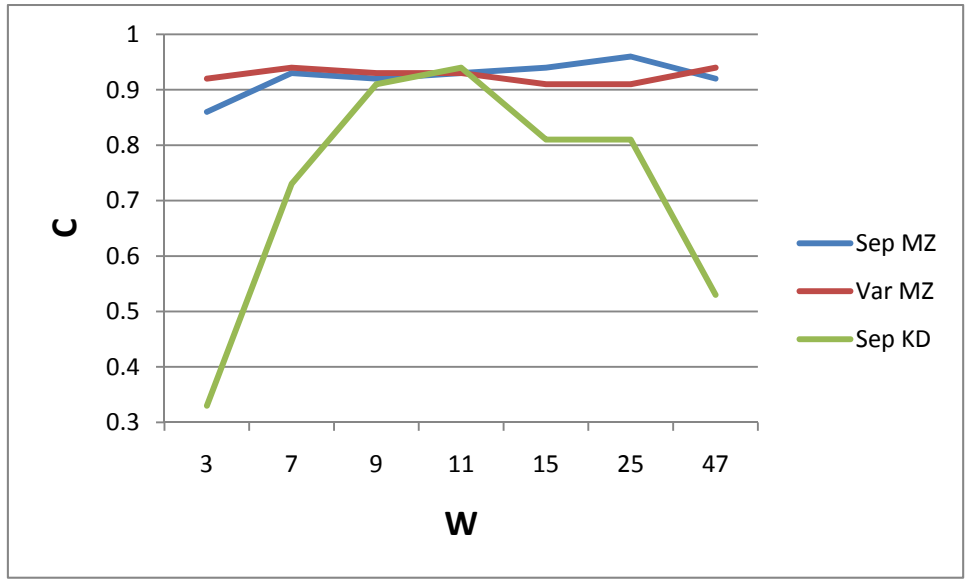

Fig. 2.